\documentclass[11pt]{article}
	
	\newcommand{\blind}{0}
	
    \makeatletter
    \renewcommand\section{\@startsection {section}{1}{\z@}%
                                       {-3.5ex \@plus -1ex \@minus -.2ex}%
                                       {2.3ex \@plus.2ex}%
                                       {\normalfont\fontfamily{phv}\fontsize{16}{19}\bfseries}}
    \renewcommand\subsection{\@startsection{subsection}{2}{\z@}%
                                         {-3.25ex\@plus -1ex \@minus -.2ex}%
                                         {1.5ex \@plus .2ex}%
                                         {\normalfont\fontfamily{phv}\fontsize{14}{17}\bfseries}}
    \renewcommand\subsubsection{\@startsection{subsubsection}{3}{\z@}%
                                        {-3.25ex\@plus -1ex \@minus -.2ex}%
                                         {1.5ex \@plus .2ex}%
                                         {\normalfont\normalsize\fontfamily{phv}\fontsize{14}{17}\selectfont}}
    \makeatother
	\usepackage{amsmath}
	\usepackage{amsthm}
    \usepackage{amssymb}
    \usepackage{braket}
    \usepackage{graphicx}
\graphicspath{{images}}
\usepackage{wrapfig}
	\usepackage{graphicx}
	\usepackage{enumerate}
	\usepackage{url} % not crucial - just used below for the URL
	\usepackage{xcolor}
\begin{document}
\def\spacingset#1{\renewcommand{\baselinestretch}
			{#1}\small\normalsize} \spacingset{1}{
			\title{\bf Eternal inflation and collapse theories}
			\author{R.L. Lechuga$^a$ and D. Sudarsky$^b$ \\
			\textit{Instituto de Ciencias Nucleares,}\\
             \textit{Universidad Nacional Autónoma de México,}\\
            \textit{ A.P. 70-543, Mexico City 04510.}}
			\date{}
			\maketitle
}
		
		\if1\blind
		{

            \title{\bf \emph{IISE Transactions} \LaTeX \ Template}
			\author{Author information is purposely removed for double-blind review}
			
\bigskip
			\bigskip
			\bigskip
			\begin{center}
				{\LARGE\bf \emph{IISE Transactions} \LaTeX \ Template}
			\end{center}
			\medskip
		} \fi
		\bigskip	
	\begin{abstract}

The  eternal inflation problem  continues to be  considered   one of standard's  cosmology most  serious  shortcomings. This arises when one  considers the effects  of   ``quantum fluctuations'' on the zero mode of  inflaton field during  a Hubble time in the inflationary  epoch. 
In the slow-roll regime it  is  quite  clear that  such  quantum   fluctuations could  dwarf  the   classical  rolling  down of the inflaton,  and  with  overwhelming probability  this  prevents inflation from  ever ending. When one recognizes that quantum fluctuations can not  be   taken as  synonymous  of  stochastic  fluctuations,  but rather  intrinsic  levels of indefiniteness in the quantities in question,  one concludes  that  the eternal inflation problem   simply does not exist. However, the same  argument  would serve to invalidate  the  account   for  the  generation of  the   primordial seeds  of  cosmic   structure as has  been  amply  discussed  elsewhere \cite{hanno, pearle, shortcomings}.
%{In order  to the   inflationary   account for the emergence of the seeds of cosmic structure to  work, and  in particular}
 In   order   to  do address  that issue,  one must  explain the    breaking of  homogeneity and isotropy  of the situation prevailing during the early inflationary epoch (at both the   quantum  and classical levels  of the description).  For that  one  needs  to  rely on   some  additional element, beyond  those present  in the traditional  treatments. The  so called   spontaneous  collapse theories offer  a viable  candidate  for  that element, namely the    stochastic  and  spontaneous  state  reduction characteristic of those proposals possesses  the basic features  to break those  symmetries.  In  fact, a version of the  CSL theory adapted to  the cosmological context  has been shown to offer   a  satisfactory  account    for the origin the seeds of cosmic structure  with  an adequate  power spectrum \cite{pearle}, and  will  serve  as the  basis of our analysis. However,  once  such  stochastic collapse  is introduced  into the theoretical framework the eternal inflation problem has the potential   reappear. 
In this manuscript we explore  those  issues  in detail    and  discuss  an  avenue  that   seems to  allow  for  a  satisfactory  account  for the generation of the   primordial  inhomogeneities and  anisotropies   while freeing the theory  from the     eternal inflation  problem.

%A common mistake in the standard analysis is the confusion between quantum fluctuations with stochastic fluctuations. It is emphasized that quantum fluctuations means uncertainties in the quantities and it lack of stochastic behaviour.

%As shown before in some references for explaining the emergence of the seeds of cosmic structure it is necessary  to incorporate an extra element to give account to the symmetry breaking vaccum.  spontaneous localization, a modification in standar quantum mechanics which introduces a non-unitary component of evolution of the field. This leads again the possibility of having eternal inflation, we point out that zero mode of inflation field is position independent then, it seems to end the problem but there still the possibility due to the others modes, whose physical wavelenght is indistinguishable to zero mode,  contribute to the problem. In this work we show that even considering those modes the problem of eternal inflation does not exist. 

%To solve this problem, we emphazise the diferent between: stochastic; space-time and quantum fluctuations and the position independence in the zero moods field.%

	\end{abstract}
			
	\noindent%

	%\newpage
	\spacingset{1.5} % DON'T change the spacing!

\section{Introduction} \label{s:intro}
%The measurment problem frecuently is considered as a problem whose only character is philosophic and it don not concern to physicist. Nevertheless has been showed in some physics areas how which problem become relevant, one of them is the cosmology context since for account of the emergence seeds structure is necessary incorporate some adjustment to standart quantum theory.\\The eternal inflation problem is taken to  be such a  serious  problem for the   inflationary cosmology  paradigm, that even   some  the  most   enthusiastic  early advocates of inflation have, nowadays, 
The eternal inflation problem is taken to  be such a  serious  problem for the   inflationary cosmology  paradigm, that even   some  the  most   enthusiastic  early advocates of inflation have, nowadays, become  some  of  the theory's  strongest detractors \cite{Ijjas:2013vea, Ijjas:2014nta,Ijjas:2018qbo}\footnote{These considerations, however, are subject to debate connected   to  the absence of a natural measure in General Relativity (GR) that can be used to estimate probabilities within the framework of GR \cite{Linde:2007nm}. }. According to \cite{linde} the quantum treatment of the inflaton  gives rise to the emergence of the eternal inflation problem due to 
{%the problem of eternal inflation arises from considering that  the inflation field requires a quantum treatment and therefore there are} 
``quantum fluctuations'' that  must  be  taken into  account  on top  of the  classical  slow roll  down  of the inflaton field  due to the combined  effect of  the inflationary potential and the  {\it  effective friction}  associated with the  rapid cosmic   expansion.
 The  argument  is  essentially the following:  consider the classical  change in the inflaton field,  $  \delta_{Class}  \phi $, as  it  ``rolls  down  the  potential"  during a characteristic period of cosmic  time $ \delta  t $, and  compare  it to the  magnitude  of the   quantum fluctuations   that can be expected  during that period  $  \delta_{Quant}  \phi $.  If the  former is  larger  than the later, then  the inflaton will evolve  towards the potential's minimum, and inflation  would  eventually end  (after a  brief reheating period).  However  if the  latter  is  larger  than the former,  and considering  the stochastic nature of the  latter, one can expect  that while    in certain   regions  the overall change   would bring  the inflaton  closer towards the potential minimum,  in  other  regions the  net change would  be in  a  direction of  an increasing value of the inflaton potential.  The crucial point, however, is that regions characterized by a higher value of inflaton potential will undergo  greater level of expansion  (as a result of being dominated by an larger effective cosmological constant) compared to regions with a smaller potential value. Thus, over time, the universe  would  end up dominated  by regions   with larger values of the inflaton potential  where the   stochastic  fluctuation increased the potential. That is  a  situation  where it would  be overwhelmingly likely for any  given  region of the universe to  be one where the  potential has  always increased.  Thus,  except perhaps   for  regions of measure that tend to zero,  we are faced  with  a runaway regime   of ever   increasing  inflaton  potential  and   never-ending accelerated   expansion  (not to  be confused  with much slower accelerated  expansion that    our universe is  currently undergoing).
 %\footnote{Actually, we know that $\frac{H_{T}^2}{H_{I}^2}\approx\left(\frac{M_{P}}{M_{GUT}}\right)^4\times10^{-120}\approx 10^{-104}$, where $M_{P}$ and $M_{GUT}$ denotes the Plank and Grand unification theories mass, respectively.}). 
 The detailed analysis that have been done concerning    this  issue conclude  that, quite generally \cite{cosmo}, the  result is  that   $ \delta_{Quant}  \phi  >>  \delta_{Class}  \phi $, indicating the  prevalence of {\it eternal  inflation}  among the  inflationary models that have been  studied \cite{Ijjas:2013vea, Baumann:2014nda, Guth:2007ng, Guth:2013epa, Boddy:2016zkn}\footnote{
 Eternal inflation has  been  atributed  mainly  to two   factors. One of them  \cite{Barenboim:2016mmw} is the extended duration  the inflaton can persist in a given  region when  the  net  displacement of the inflaton takes it  away from the zero of the potential. Another  factor \cite{Guth:2007ng} %(in which we will have an in-depth discussion in the following) is due
  is that generically,   when the field ``rolls backward"  the value of  potential  increases,    and  so does the expansion rate, and  as time passes, regions with a higher expansion rate come to predominate.}. 
 
   However, a close  examination of the  arguments above, unveils that several delicate issues are at play, and that several assumptions have  been implicitly employed, assumptions that are not only far from inescapable,  but  often  an problematic and  not really  appropriate.   
  
  Let start by noting that what drives the accelerated expansion  during inflaton it is the zero mode of the inflaton field, the only mode that is supposed to be highly excited and sharply peaked, to the point that it is treated  in classical terms in most works  on inflation. \footnote{ See \cite{diez} for a an example  where the  zero mode  is given a full  quantum treatment.}
  The  other modes of the field are supposed to be  in the  vacuum state ( more precisely a suitable adiabatic vacuum state),  often taken as the Bunch  Davies  vacuum  state,  (in part  as a result the extremely  rapid   cosmic  expansion). The potential is then, up to small corrections, essentially determined by the value of the  zero mode of the inflaton. On the other hand,  as the zero mode is, by definition,   invariant under spatial  translations,  any  variation in the zero mode can not result in  differences in what takes place in  different  locations, so the argument that the fluctuations  would lead to increases or decreases in the value of the  potential in different  regions  can not possibly refer to anything associated  with the zero mode.   
  
 Next, let us point out that the  identification of quantum  fluctuations with stochastic  fluctuations  is not something that can be justified   in the context of   standard versions of quantum theory. The   ground state of  the hydrogen  atom  does not reflect -- according to  the textbook  versions of quantum  mechanics--  an  electron moving  stochastically  within the region where   the  value of   wave function is large.   That type  of  picture  might  arise in   some  modified versions of the theory  but  it is  certainly not  what  the Copenhaguen  or von Neumann  interpretations say.  In fact,  standard  versions of quantum theory contain absolutely no  stochastic elements  except for  those  that  occur in  a  association with the  reduction postulate. That is,  stochasticity makes its only appearance in the  theory  in connection  with the measurement process, for otherwise the evolution is provided  by  the  Schr\"odinger equation which is completely  deterministic regarding the wave function. Moreover as the theory  explicitly states that the  wave  function provides a complete  description of the physical situation, there  is nothing else  to which  to associate   a  stocasticity. It is  thus clear  there is nothing that can possibly  be said to behave stochastically unless one  makes use of  the reduction postulate, which  of course  brings up the notion of observers or devices,  which  as  stressed,  elsewhere are hared  to  make sense of  the cosmological context (See for instance \cite{bell2004quantum}).
 
 This last point in fact  takes  straight us into  the   delicate questions  surrounding  the so called  measurement problem  in quantum mechanics \cite{Maudlin, RPenroseUyR},   and  its extensions  into  the even more complex realm of quantum field theory \cite{Bellviej} in   curved  space-time \cite{diez}.  It is not the intention of the present  work  to  enter  deeply into this  discussion, as  that has been the  object of  extensive discussions, and in  fact various other  works have focused on the  specific form the problem takes in the contexts of inflationary cosmology \cite{hanno, ward}. Lets  just   note that  as discussed already  in \cite{bell1},  the  cosmology  setting is  one where the  conceptual problems that besiege quantum theory  are  exacerbated  by the fact that one is  dealing  with a situation where  no  measuring apparatuses or observers  can be call upon to justify the use of the usual reduction postulate. When considering,  in  particular,  the inflationary account of the emergence of the seeds of cosmic  structure, one  faces the further obstacle  that, the  evolution of such structure, leading the formation of galaxies, stars, black holes, etc,  is a prerequisite for the  emergence of   observers  and  measuring devices. When  attempting to face such difficulties, and more  generally the    measurement problem in quantum theory,  one can be helped  by the  work \cite{Maudlin} which helps to  classify  the  viable logical  paths  open  to deal with the measurement problem, as: i)   hidden variable theories\footnote{Which  as per  the experimentally  verified   violation  of   
 Bell's Inequalities ought to be non-local (as is the case with the de-Broglie Bohm approach \cite{bell1}).},  ii) spontaneous reduction theories \cite{ghirardi, Bedingham1, Bedingham2, Tumulkagrw}, or iii) many worlds type  approaches, which includes the   bold  attempt in \cite{Hartle:2016tpo, Hartle:2015gfu, Hartle},  which  however seems to  suffer  from   very  serious  problems \cite{Okon:2013kc}. 
 The difficulties  regarding  of   inflationary  account of  the emergence  of  the  primordial  inhomogeneities  have   been  the  subject of  explorations  based  on all those  approaches: The works  \cite{Pinto-Neto, ward}  are examples of  i),   \cite{pearle, Martin:2021lje, Pearle2014tda} are representative    of   ii),  while  an implicit   acceptance   of iii)   seems to underlay  most of the  works on the subject that do not explicitly acknowledge the issue. There have  been various  works  focused  on the problem  (such as \cite{Kiefer, Ashtekar},  while  trying to avoid  taking a definite  posture  regarding  options  i)  ii)  and iii),   have  been found to suffer  from serious  shortcomings \cite{Debunking, Sudarsky:2009za}). 
 
 The fact is that, having argued that in  standard  versions of quantum theory there is nothing  stochastic unless  a  measurement is involved, and having noted that in the cosmological  situation at hand there is no   justification for  calling   on  the reduction postulate,  we  recognize that   something must be added into the picture in order to break the homogeneity and isotropy  reflected both in the aspects treated classically,  (often  just the  background) and   in  vacuum quantum state characterizing the conditions of the degrees of freedom that  are given a quantum treatment. In other words,  within  the  inflationary context all modes of the quantum fields involved   other than the  zero mode  of the inflaton  are supposed to be in the adiabatic   vacuum  and   such   state is completely homogeneous  and isotropic, despite the  quantum  uncertainties  for various  quantities\footnote{At this point we   might note  that  just as   the  adiabatic  vacuum state of  such quantum fields  is characterized  by   typical  uncertainties,  so   is the  ordinary Minkowski  vacuum  and that such   uncertainties   in no  way affect the  complete invariance of the Minkowski vacuum under the full  Poincare group.},  so  whatever element   we  add  to the theoretical picture,  it should  be  able to   account for the  breakdown   of such  symmetries  during the subsequent  evolution.

Here we will focus on a treatments based on  spontaneous  collapse,  or dynamical  reduction  theories     
which  are based on the  idea  of unifying the  unitary and deterministic evolution   provided by the Schrödinger  equation and the indeterministic    changes  normally  associated measurement situations  by adding to the former  suitable     stochastic  terms.  That modified  evolution is taken to be  universally valid, requiring no   
{\it a priori}  distinction between    situations involving measurements and   those   that do not.   When applied to the    inflationary cosmology  setting it   offers  a novel account to the   breaking the homogeneity  and  isotropy of the   adiabatic  vacuum. 
   In  that context  the evolution of the quantum state  is no longer  fully deterministic despite the absence of measurements  and  observers,  and  in fact,  stochastic elements  now do appear in the  dynamics \cite{hanno, pearle1990toward},   which   can    help to account for the emergence of the  primordial seeds of cosmic  structure \cite{hanno}, however,   at the same time, those  stochastic element in the   dynamics   resurrect the specter of eternal inflation.

  In this  work,   we  will consider a treatment   based  on  the Continuous  Spontaneous Localization theory \cite{pearle1990toward},  and adapted  in a particular manner \cite{pearle}  to the  inflationary cosmology  setting, and   which, offers a  successful treatment 
  of the problem including the  possibility  to  account  for the an adequate  primordial power spectrum  of density fluctuations \footnote{ We  note that  the  predictions  for the   power spectrum  of tensor modes   in this   approach    differs  substantially from the  standard accounts \cite{Palermo:2022dim, Susana}.}.
  
%    The particular approach  will  study  takes  the  quantum field  itself  to   play  the role  of the   collapse operator\footnote{ The  operator  that accompanies the  new stochastic terms that the  theory adds to  the  Schr\"odinger evolution.}. 

 % In that case, as 
%   discussed  in \cite{pearle} the  collapse rate 
%     $\lambda$ must,  as indicated  on  dimensional  grounds    
%     and  in  order to  generate the essentially  scale invariant  primordial  spectrum,  depend on the wavelength of the  collapsing mode\footnote{This, as  shown in  \cite{Leon:2017sru}  
%     can    be implemented  alternatively by  choosing as collapse operator a suitable  expression involving  derivatives of the  inflaton field.}.   Concretely, what  one   finds 
%     $\lambda=  \lambda (k) =  k\tilde\lambda $,  where $ k $ is the 
%       mode's co-moving  wave number  and   $\tilde  \lambda $ is an universal parameter of order $\lambda \approx 10^{-5}MpC^{-1}$ (which exhibits favorable comparisons with the range of values typically associated with the collapse rate in the non-relativistic many-particle quantum mechanical framework for which the theory was initially developed to address).

The   study  carried out in   \cite{pearle} considered two  scenarios 
for the choice of the   collapse operator\footnote{ The  operator  that accompanies the  new stochastic terms that the  theory adds to  the  Schr\"odinger evolution.}, the field  itself and  its momentum conjugate.
 The conclusions indicate that  in the first case,   in  order to  generate an essentially  scale invariant  primordial  spectrum, the collapse rate 
     $\lambda$ must,   as is  also  suggested   on  dimensional  grounds,   depend on the wavelength of the  collapsing mode
     %\footnote{This, as  shown in  \cite{Leon:2017sru}  
    % can    be implemented  alternatively by  choosing as collapse operator a suitable  expression involving  derivatives of the  inflaton field.}.   Concretely, what  one   finds 
     according to 
     $\lambda=  \lambda (k) =  k\tilde\lambda $,  where $k$ is the 
       mode's co-moving  wave number  and   $\tilde{\lambda}$ is an universal parameter of order $ \sim 10^{-5}Mpc^{-1}$ (which   as    noted  in \cite{pearle} exhibits favorable comparisons with the range of values typically associated with the collapse rate in the non-relativistic many-particle quantum mechanical framework for which the theory was initially developed to address).

       For the  second case,  and   again as  suggested  by dimensional considerations,   it turns  out that 
    in  order to  generate the essentially  scale invariant  primordial  spectrum,   the collapse rate   must depend on the wavelength of the  collapsing mode
  according to 
     $\lambda=  \lambda (k) =  \bar\lambda/k $,  where $ k $ is the 
       mode's co-moving  wave number  and   $\bar  \lambda $ is again  universal parameter. In fact a related  study   addressing the  issue  of  eternal inflation  in the context of spontaneous  collapse theories  have been carried out  in   \cite{Leon:2017sru}  focusing  explicitly in this  second  scenario. 
       
       I the present work we   will  consider the first   scenario,  which we find  more natural  given  the  lack of  pathological  behavior of the  collapse rate  in the vicinity on  the  zero mode  ( i.e.  as  $ k\to 0$).

The  conclusions for the case we examine, strongly suggests that the  zero mode  should  not  be affected by the spontaneous   collapse dynamics, and  thus  would not  suffer  from stochastic  fluctuations  therefore fully evading    the  eternal inflation   problem. 
Unfortunately things are not so simple,   and  one must  explore  the  effect  of the modes  other  than the  zero mode  but  having   sufficiently long   wave  length to behave ``effectively" as the zero mode. 
The detail   analysis of that question  is one of the main objectives of the present  work.

%If one choose the $k$ modes such that $a(t)\frac{2\pi}{k}>>H_{DP}$, where $H_{DP}$ is the particle horizon, it is possible to conclude that eternal inflation condition leads $\frac{\frac{d}{d\eta}\sqrt{\overline{\braket{\delta \phi}}}d\eta}{\dot{\phi_{0}}(\eta)d\eta} <<1$. In other words, it is possible to show that eternal inflation never appear when collapse theories and semi-classical gravity approach are incorporated at the standard treatment.   

%\section{The emergence seeds results by collapse theories} \label{s:methods}

%I will write the main results for attempt to problem of emergence seeds.
%It is important: The master equation of CSL, The result for the spectrum of $\overline{{\langle X^2 \rangle} }$ and $\overline{{\langle P^2 \rangle} }$ as function of $\lambda$. 

\section{Inflation and cosmic structure}

The majority of contemporary cosmological models incorporate an  early   inflationary  epoch characterized by an accelerated expansion, as an  integral  part.
One of its most attractive aspects is it claim  to offer a natural explanation for the emergence of the primordial seeds of  cosmic structure as a result  of quantum fluctuations. In this section we present   a brief description of the  standard  accounts of that.

The model is described by the action,
\begin{equation}\label{acc}
    S=\int d^{4}x\sqrt{-g} \left\lbrace  \frac{R}{16\pi G}  -\frac{1}{2}(\nabla_{a}\phi\nabla_{b}\phi g^{ab} + V(\phi) \right\rbrace,
\end{equation}
 with  the matter sector  characterized by a scalar field $\phi$, named the inflaton. This work is focuses on the ``chaotic inflation model'' associated  with   the  quadratic  potential,  $V(\phi)=\frac{1}{2}m^{2}\phi^{2}$ as a representative example\footnote{According to \cite{Planck:2015fie} the inflationary models with $V(\phi)\propto \phi^{2}$ are disfavoured, however as shown  in \cite{Piccirilli:2017mto, Susana} when   the analyses  is made using collapse theories in a semi-classical context instead of the standard treatment such potentials are not problematic.}. The background it is given by the flat Friedman-Lemaître-Robertson-Walker (FLRW) space-time, whose line element is expressed in conformal time, $\eta$, is:
\begin{equation} \label{FRWC}
    ds^{2}=a^{2}(\eta)[-d\eta^{2}+\delta_{ij}dx^{i}dx^{j}].
\end{equation}
The Einstein field equations are
\begin{equation}
    G_{ab}=8\pi G T_{ab},
\end{equation}
with  the energy momentum tensor for the scalar  field   given by,
\begin{equation}
T_{ab}=\nabla_{a}\phi \nabla_{b}   \phi -\frac{1}{2}g_{ab}(\nabla_{c}\phi \nabla^{c}\phi+m^{2}\phi^{2}).
\end{equation}

%It is well known that there are numerous attempts to relate quantum theory and general relativity. Nevertheless so far  there is no fully adequate theory of quantum gravity. \\

%so far that in physics lack a quantum theory which incorporates gravity....STILL REMAIN LACK REVIEW OF SEMICLASICAL GRAVITY
%Choose a basis coordinates the relevant E eq are 
The relevant Einstein equations are:
\begin{equation}
   3\mathcal{H}(\eta)^2=4\pi G \left[\left(\frac{d\phi_{0}}{d\eta} \right)^{2}+2a(\eta)^{2}V(\phi_{0}) \right],
\end{equation}
and the Klein-Gordon equation is,
\begin{equation}\label{KGn}
    \frac{d^{2}\phi_{0}}{d\eta^{2}}+2\mathcal{H}(\eta)\frac{d\phi_{0}}{d\eta}+a(\eta)^{2}\frac{\partial V}{\partial \phi_{0}}=0.
\end{equation}

The setting  in  which  the process of  interest takes place  is  taken to be   
 that  of the so call ``slow-roll regime",  corresponding  to  the   well known slow-roll condition, 
\begin{equation}\label{slowr-cond}
    \frac{d^{2}\phi_{0}}{d\eta ^{2}}=\mathcal{H}(\eta)\frac{d\phi_{0}}{d\eta} ,
\end{equation}
where $\mathcal{H}(\eta)=a(\eta)H_{I}$ and $H_{I}=\sqrt{\frac{8\pi G V}{3}}$.  Combining (\ref{slowr-cond}) with (\ref{KGn}) one  obtains; 
\begin{equation}
  \frac{d\phi_{0}}{d\eta}=-\frac{a(\eta) m^{2}\phi_{0}}{3H_{I}}.
\end{equation}
Recall  that  to a good  approximation  we  can take $a(\eta)=\frac{-1}{H_{I}\eta}$, and set  the inflationary period to correspond to  $-\tau <\eta <\eta_{f}$ (where inflation  takes place, begins and ends) and $\eta_{f}<0$. 
The above   expression  for  the scale factor taken  $ H_I$ as  a  constant,  is only an approximation. Alternatively  we can take  $H_{I} $ to have a  small time dependence  due to the  slow  but non-vanishing   changes  in $V$  as inflation proceeds.  %For simplicity, we  will ignore this   aspect in the present  treatment as it implies  no  fundamental  changes in   the overall   analysis. 
As is  often done, we  take the expression for $a(\eta)$ to  be $a(\eta)=\left(\frac{-1}{H_{I}\eta}\right)^{1+\epsilon}$, the factor is written in terms of slow roll parameter and  $\epsilon=1-\frac{\mathcal{H}'}{\mathcal{H}^{2}}$.

In the standard treatment employed to account for the emergence of  the first cosmic seeds structure one  considers  perturbations for both, the inflaton  field $\phi=\phi_{0}+\delta \phi(\eta,\vec{x})$, and takes the metric  which (using  a specific gauge and ignoring  vectorial and tensorial perturbations)   to have the form,
\begin{equation}\label{perturbada}
    ds^{2}=a^{2}(\eta)[-(1+2\psi(\eta, \vec{x}))d\eta^2+(1-2\psi(\eta, \vec{x}))\delta_{ij}dx^{i}dx^{j}].
\end{equation}
It is customary  to define a new  variable  linking $\psi(\eta, \vec{x})$ with $\delta \phi(\eta, \vec{x})$,
\begin{equation}
    v\equiv \left( \delta\phi + \frac{\dot{\phi_{0}}}{\mathcal{H}}\phi\right).
\end{equation}
Then, one  proceeds to quantize this  variable constructing the Fock-Hilbert space and  to  write the corresponding  field  in terms of creation and anihilation operators 
\begin{equation}
    \hat{v}(x,\eta)=\sum_{\vec{k}}\left(\hat{a}_{\vec{k}}v_{k}(\eta)\exp{(i\vec{k}\cdot\vec{x})}+\hat{a}_{\vec{k}}^\dagger v_{k}^{*}(\eta)\exp{(-i\vec{k}\cdot\vec{x})} \right).
\end{equation}
The modes are chosen  in such a way  that, at  $\eta \rightarrow -\infty$ behave as  the  positive  frequency  standard modes  in  Minkowski space-time. This  corresponds  to the selection of a particular vacuum state $\ket{0}$,  via the condition  $\hat{a}_{\vec{k}} \ket{0}=0$, namely   the ``Bunch Davies'' state in  de-Sitter space-time   and  to the so  called  ``adiabatic vacuum''  in  the  case of  our  slow rolling  inflationary model.

One is interested in the   characterization   primordial seeds of  structure as   given by  the relative density fluctuation  during inflationary regime $\delta (\eta, \vec{x})=\frac{\delta \rho(\eta, \vec{x})}{\overline{\rho(\eta)}}$, where $\overline{\rho(\eta)}$ is the spatial average of the universe's density, $\rho(\eta, \vec{x})$, $\delta\rho(\eta,\vec{x})\equiv \rho(\eta,\vec{x})-\overline{\rho(\eta)}$.
The power spectrum characterizing   those primordial  inhomogeneities  is taken to be characterized statistically by the correlation:
\begin{equation}
    \overline{\delta(\eta, \vec{x})\delta(\eta, \vec{y})}=\frac{1}{(2\pi)^3}\int d^{3}k P_{\delta}(\eta,k)\exp{(i\vec{k}\cdot(\vec{x}-\vec{y}))}.
\end{equation}
 where the quantity of interest  is $ P_{\delta}(\eta,k)$  known  as the power spectrum  of density fluctuations.

The  analysis  proceeds  by taken the latter  as   represented  the two point function: 
\begin{equation}
    \bra{0}\hat{v}(x,\eta)\hat{v}(y,\eta)\ket{0}.
\end{equation}
  Thus from the last expression the power spectrum, is   extracted through:
\begin{equation}
    \bra{0}\hat{v}(x,\eta)\hat{v}(y,\eta)\ket{0}=  \int d^{3}k \exp{(i\vec{k}\cdot(\vec{x}-\vec{y}))}P(k).
\end{equation}
 The result is essentially  $P(k)=Ck^{-3}$ (up to  a  small correction  associated  with the  slow roll  parameter  $\epsilon $  which we will  ignore  henceforth).
%That  characterization  is identified  with the  primordial seeds of  structure as   given by  the relative density fluctuation  $\delta (\eta, \vec{x})=\frac{\delta \rho(\eta, \vec{x})}{\overline{\rho(\eta)}}$, where $\overline{\rho(\eta)}$ is the spatial average of the universe's density, $\rho(\eta, \vec{x})$, $\delta\rho(\eta,\vec{x})\equiv \rho(\eta,\vec{x})-\overline{\rho(\eta)}$.
In other words  one  argues that the ``quantum fluctuations''  seed the   primordial   inhomogeneites and  anisotropies that  eventually evolve into all cosmic  structure present in our universe. 
%The power spectrum characterizing   those primordial  inhomogeneities  is taken to be characterized statically by the correlation:
%\begin{equation}
%    \overline{\delta(\eta, \vec{x})\delta(\eta, \vec{y})}=\frac{1}{(2\pi)^3}\int d^{3}k P_{\delta}(\eta,k)\exp{i\vec{k}\cdot(\vec{x}-\vec{y})}.
%\end{equation}
This result is considered as one of the biggest successes of the inflationary  model.  %epoch, since it is customary interpret it as the result to lead the formation of first seeds structure.

 At this point it is worthwhile emphasizing  that  the  delicate issue of the   meaning of  ``quantum fluctuations''  in this context. Before  doing  so, it is important to consider   the differences between the three following  notions:

\textbf{Space-time fluctuations:} This kind of fluctuations describe a unique and wide object on which a  locally defined quantity  changes from one point to another, an example,  the temperature on each part of  fluid at a  definite  time.\\
\textbf{Statistical fluctuations:} Those are associated to an ensemble, each  element of which  having   definite  value of the quantity of interest. %a weight, $\omega_{i}$.
  This fluctuations describes the variations of such  value  among  the  various elements of the  ensemble.\\
  \textbf{Quantum fluctuations:} 
This type of fluctuations usually describe indeterminacies of    some  operator  when the system in question is  in a given quantum state. An example  is provided  by a  particle that  does not  have a definite momentum or position, such as,    say,  a  harmonic oscillator in its  ground state.\\
One must  note that in the previous treatment the last three notions are taken as synonymous.  In  particular many cosmologists  treat  ``quantum fluctuations'', quite  naturally  as   if   they  spacetime  or  statistical fluctuations. It is well known in quantum theory that quantum uncertainties play a significant role. In particular they characterize the dispersion of outcomes observed when measuring an observable on identical systems. According to the standard theory when a measurement is performed, the state of the system transitions from the  state that in general does not have a well-defined value of the observable in question to a new state that possesses a well-defined value (or at least a better-defined value, if the measurement is not  a perfectly accurate one).

However as noted in  above (and  amply discussed elsewhere)  quantum fluctuations describe only indeterminacies and lack statistical behaviour.
%Morover the  quentum fluctuationss  do nothing to change the homogeneity and isotropy of the  quentum state. 
Thus, the  analysis  presented  above,  without some  additional input  is simply  unjustified  as  an  account of the  emergence of structure. In particular  it  contains no  element    accounting  for the  beak down of the  complete homogeneity and isotropy of  the situation,  because  vacuum state possesses   those   symmetries  and there  is  nothing in the  account that  would  justify  arguing that those are broken. Inspired in this problem, in reference \cite{pearle, Piccirilli:2017mto}  (following    various  previous   works  based on  simplistic versions  incorporating  some   kind of   spontaneous state  reduction \cite{hanno})    consider a modified approach with an specific proposal: incorporate continuous spontaneous localization, a modified quantum theory, designed   to solve the measurement problem in  ordinary  quantum  mechanics,  and  adapt it to the  context of semi-classical gravity. The following section describes the relevant results of   \cite{pearle},    the work  on  which the   subsequent  analysis of the    eternal inflation problem will be based.

\section{Seeds of cosmic structure and collapse theories}
\subsection{Considerations   about  the  treatment  to  be adopted}

 Here  we   will offer a brief review  of the  work in  \cite{pearle}  presents  an approach to account  for  the generation  of primordial anisotropies and inhomogeneities during the infationary period that is  explicit and clear regarding the   mechanism  by  which the  symmetries of the vacuum state  are broken. Before  discussing  that  it  its  worthwhile  to touch on  another delicate aspect in the standard treatment in order to  clarify the steps taken in this work. Specifically, we address the challenge of working at the interface between quantum theory and gravitation, which is further complicated by the absence of a fully workable  theory of quantum gravity (although certainly substantial efforts  have  been  made  in   number of approaches  to the subject). 

   The point is that  while it is    common to think that the   difficult questions  involving    quantum gravity refer exclusively the   Plank regime,  problematic issues emerge in a much broader set of circumstances. One  aspect  of the problem can  already  be seen to  occur  in  simple  situations  such as that  discussed in  \cite{donpage} where  an actual experiment  involving  a  gravitating body that is  arranged to go   into    macroscopic  superposition  of two  different  positions,  and employed arguable   to illustrate the invariability of the semi-classical General Relativity (GR).  More specifically the  claim is  \cite{DanielGhirar}   that  1) If there are no quantum collapses, then semi-classical GR conflicts with the results of their experiment. 2) If there are quantum collapses, or reductions of the wave  function,  then semi-classical GR equations are internally inconsistent. The last point  refers to the  fact that the left side of Einstein's equation, $G_{\mu \nu}=8\pi \braket{\hat{T}_{\mu \nu}}$,  is  divergence free as a result of Bianchi's identities while that  the expectation value of the energy momentum tensor, will not be  divergence free if a  collapse  takes place. 
    However,  as we noted   we  do not consider that  we  have a really solid  alternative    and thus    we are  inclined to use    the semi-classical GR  and  consider  it, not as a  complete  and    consistent theory on its  own ,  but   as an approximated description with limited domain of applicability. 
    On the  other hand  we must  acknowledge that other approaches have been used to incorporate   the collapse theories in considering  inflationary cosmology \cite{shadow} \cite{singh}\footnote{ We are  not truly persuaded by  such  approaches,  in part  because  we share the concern:   $\ll$The breakup of the metric into a background metric which is treated classically and a dynamical field $\gamma_{ab}$, which is quantized, is unnatural from the viewpoint of classical general relativity. Furthermore, the perturbation theory one obtains from this approach will, in each order, satisfy causality conditions with respect to the background metric $\eta_{ab}$ rather than the true metric $g_{ab}$
$\gg$, 
see in page 384, second paragraph of \cite{ward}.}.   For truly fundamental  characterization, we   think a  full theory  of quantum  gravity would be required. That is  we take the view  that  a  theory  with  spontaneous quantum collapses not only    offers  a  viable  resolution of measurement problem   but  that  it can  provide  an approximate effective  description of   situations  involving gravitation  quantum  matter, despite the fact that   during the  collapse  events the equations can not be valid and require a delicate handling.
    
Therefore, we will consider semi-classical gravity as an analogous  to  say the   Navier  Stokes equations   in   hydrodynamics   which  although  clearly not fundamental equations failing to reflect in full the   atomic  molecular nature of  fluids,  often provides a rather  good  characterization of their  effective behavior.    Such  descriptions   will often fail when the  behavior  of fundamental degrees  of freedom of the fluid  become   sufficiently excited   for the approximations  involved in the hydrodynamic  description cease to be valid (say    at the  onset  of turbulence).    Similarly  we assume  that  semi-classical GR equations are valid before and after a spontaneous collapse, but not at the time in which it is occurring.  A formalism  describing the  treatment of that  kind of situation  have  been  presented in  \cite{diez}   and further  developed in other works  \cite{pearle, Juarez-Aubry:2022qdp} . The   essence of  that formalism  relies on  precise  descriptions for the situations just before and just after a collapse,  which  are  subject ``gluing procedure" inspired  on  the treatment of \cite{Israel}.
The precise   characterization of the situations  before and after  a  collapse are  to be given   in terms  of Semi-classical Self-consistent Configuration (SSC), defined    as  follows:

The  set 
\begin{equation}
   \{ g_{\mu\nu}(x), \hat{\varphi}(x), \hat{\pi}(x), \mathcal{H}, \ket{\xi} \in \mathcal{H}\}.
\end{equation}
It is said  to represents a SSC iff $\hat{\varphi}(x)$, $\hat{\pi}(x)$ 
correspond to  suitable   field  and  momentum conjugate operators  acting  on  the  Hilbert space $\mathcal{H}$   for a quantum field theory in the  space-time with metric  $ g_{\mu\nu}(x)$, and  where $\ket{\xi}$  is  a  state in the  Hilbert space  that satisfies :
\begin{equation}
    G_{\mu\nu}[g_{\mu\nu}]=8\pi G\bra{\xi}\hat{T}_{\mu \nu}[g(x),\hat{\varphi}(x),\hat{\pi}(x)] \ket{\xi},
\end{equation}
for each point $x$ in the space-time.

 Building an SSC is not trivial it is necessary give an ansatz, one must propose the appropriate metric of space-time, build quantum theory, find an appropriate state, $\ket{\xi}\in \mathcal{H}$, that is compatible with the selected configuration of space time. 

Using  the SSC approach  as  shown in  \cite{diez}  one can explicitly   describe the  the transition from an early homogeneous and isotropic  stage of the  universe   one that is inhomogeneous and anisotropic. The  additional  element  is  provided  by the spontaneous  and   stochastic   quantum collapse of the wave function of the  matter fields. 

The proposal, by incorporating the CSL theories, is to be able to explain this transition considering normal unit evolution, characteristic of standard quantum field theory but supplemented by quantum collapse, which has been suggested can somehow predict the gravitational degrees of freedom. 

In this work they consider that any state can be understood as a particular SSC, which is why the transition from one state to another, mediated by collapse, can be read as
\begin{equation}
   \textrm{SSC-I} \rightarrow  \textrm{SSC-II}.
\end{equation}
 Inspired by this work \cite{diez}, which explains in a general manner how one evolves from one hypersurface to another through a collapse, the following scheme \cite{pearle} is employed to account for the formation of primordial structure.
 
 It is  worth mentioning that recent  substantial progress haas  been achieved   in  putting those ides    in  firmer mathematical  ground \cite{ Juarez-Aubry:2019jon, Juarez-Aubry:2021abq, Juarez-Aubry:2022qdp}.
 %\cite{Tona1, Benito, Tona 2}.

%%%%%%%%%%%%%%%%%%AQUÍ SE CAMBIA %%%%%%%%%%%%%
\subsection{A revised  account  of the  emergence  the    seeds  of  cosmic structure}
The  work  \cite{pearle}  presents a consistent treatment to describe the emergence of the seeds of cosmic structure   using  an  approach that incorporates semi-classical gravity and the CSL version  of  spontaneous  collapse theories.  Most of this  section is  a recount of that  work  with small adaptations required  for our purposes. 
The staring  point is the   usual  description in the inflation regime  with its  basic dynamics  given  by the action:
\begin{equation}\label{acc}
    S=\int d^{4}x\sqrt{-g} \left\lbrace  \frac{R}{16\pi G}  -\frac{1}{2}(\nabla_{a}\phi\nabla_{b}\phi g^{ab} + V(\phi) \right\rbrace,
\end{equation}
leading to  the  field equations 
\begin{equation}
    G_{ab}=8\pi G T_{ab},
\end{equation}
with  $T_{ab}$, 
\begin{equation}
T_{ab}=\nabla_{a}\phi \nabla_{b}   \phi -\frac{1}{2}g_{ab}(\nabla_{c}\phi \nabla^{c}\phi+m^{2}\phi^{2}).
\end{equation}
Prior to conducting the analysis proceeds  by  separating the metric and scalar field into a spatially homogeneous-isotropic background and  the perturbation part describing the  possible   departures  from   the exactly  symmetric   situation,
\begin{equation} \label{FLRW-pert}
    ds^{2}=a^{2}(\eta)[-(1+2\psi)d\eta^2+(1-2\psi)\delta_{ij}dx^{i}dx^{j}],
\end{equation}
where   again one  is  working in the  Newtonian  gauge and  the vector and tonsorial perturbations have been ignored. The scalar field is, (in   following with  the semi-classical   spirit in which the  space-time metric  is given a classical treatment and  the matter fields  a full   quantum one) treated in principle    in quantum  terms  at the level of both  background  and perturbations. %although we will see that it will be equivalent to treat it in a classical way.
We  however  separate  field into  the  space independent, or   zero mode  $ \phi_{0}$,  and the rest : 
\begin{equation}
    \phi(x)=\phi_{0}+\delta \phi(\vec{x},\eta),
\end{equation}
 where  the    $\delta \phi (\vec{x}, \eta)$  represents what is normally  considered  in terms of the  inflaton field perturbation   which is the only  part that  is  usually  given a  quantum treatment. Expanding the  action  up to the second order in the   scalar   field ``field perturbations" we obtain   after  making a  convenient change of  variable $y\equiv a\phi$, 
 \begin{equation}
  \delta S^{(2)}=\frac{1}{2}\int d\eta d^{3}x (y'^{2}-(\nabla y)^{2}+\mathcal{H}y^{2}-2\mathcal{H}yy').  
 \end{equation}
 
From  this  we  obtain the corresponding  Hamiltonian, 
\begin{equation}
    \delta H^{2}=\frac{1}{2}\int dx [\pi^{2}(x)-\frac{2}{\eta}\pi(x)y(x)+(\nabla y(x))^{2}],
\end{equation}
where we have  suppressed  the  implicit  dependence on $\eta$.
In  fact it  is convenient to   work in the  Shr\"oedinger picture  and  focus  on the  behavior of   the relevant  expectation values of Fourier   decomposition   of field  and momentum  operators.
 In principle  we  should    separate each of  these  into  symmetric and anti-symmetric  parts in order to  work with bonna-fide  hermitian operators,  but as    shown in \cite{pearle}, the    results obtained  in a simplistic  analysis that  avoids that step  are   the  same  as those  found  using   more explicitly  rigorous  one.    %Subsequently, we proceed to  analyze  the behavior of each mode of the field.
Then   we  write   the field  and   conjugate momentum operators as:
\begin{equation}
   y(\vec{x}) =\frac{1}{(2\pi)^{3/2}}\int y(\vec{k})e^{i\vec{k}\cdot\vec{x}}d^3\vec{k} ,
\end{equation}
\begin{equation}
    \pi(\vec{x})=\frac{1}{(2\pi)^{3/2}}\int \pi(\vec{k})e^{i\vec{k}\cdot\vec{x}}d^3\vec{k} ,
\end{equation}
The commutation relations  are then,  
\begin{equation}
    \begin{split}
        [y(\vec{k}),\pi(\vec{k'})]&=\delta(\vec{k}-\vec{k'}),\\
        \delta H&=\frac{1}{2}\int d\vec{k} \left[\pi(\vec{k})\pi^{*}(\vec{k})-\frac{1}{\eta}[\pi^{*}(\vec{k})y(\vec{k})+\pi(\vec{k})y^{*}]+k^{2}y(\vec{k})y^{*}(\vec{k}) \right].
    \end{split}
\end{equation}
 As noted,  in order to give a quantum treatment and work with hermitian operators the description is made in terms of symmetric and anti-symmetric  components of the field   and momentum  finding   a simple collection of independent modified-harmonic oscillators. We focus our attention on one specific   Fourier mode $k$ (and treat all in a similar manner). The operators that will turn out to be convenient are introduced:
\begin{gather}\label{xx}
    \hat{X}_{k}=\sqrt{d^{3}\vec{k}}\hat{y}(\vec{k})\\
\label{pii}     \hat{P}_{k}=\sqrt{d^3\vec{k}}\hat{\pi}(\vec{k}),
\end{gather}
where $d^3k$ represents an ``infinitesimal'' volume in the space of the $k$'s around $\vec{k}$ (i.e. correspond to  $(2\pi/L)^3$ if we ``put the universe in a  box" of  size  $L $ and impose   periodic  boundary conditions).  Then, (\ref{xx}) and (\ref{pii}) satisfy the standard   discrete commutation rules $   [\hat{X}_{k},   \hat{P}_{k}]= i \delta_{k, k'} $. The hamiltonian for the mode $k$ is:
\begin{equation}
    \hat{H}_{k}=\frac{1}{2}\left[\hat{P}^{2}-\frac{1}{\eta}[\hat{P}\hat{X}+\hat{X}\hat{P}]+k^{2}\hat{X}^{2} \right].
\end{equation}
In the treatment done above, the zero mode of the field is not  included, because  we  will see it, in the   approach we  use  it   will  naturally   be unaffected  by the  modified   for which a more detailed analysis will be given in the following sections.

And noted we  will  focus the analysis on a single mode $ k$  so  in the  reminder of the paper the sub-index  $k$ will be   dropped  so,  for instance,   $\hat{X}_{k}$  and  $\hat{P}_{k}$ will be    denoted  by $\hat{X}$  and  $\hat{P}$ respectively. 

\subsubsection{Quantum collapse free scenario}

Before  considering what happens once the   CSL modifications  are  incorporated,  one  consider the  behaviour of the relevant expectation values  when  only the  standard unitary  evolution is  considered (i.e   the CSL  modifications  are turned off). The Hamiltonian of the system  is $H_{k}$,  but as  we are working in the  Schrödinger picture  we have:

\begin{equation}
    \frac{d}{d\eta} \bra{\psi, \eta}\hat{A}\ket{\psi, \eta}=-i\bra{\psi, \eta}[\hat{A}, \hat{H}]\ket{\psi, \eta},
\end{equation}
in the following, we  will be denote  $\braket{A}= \bra{\psi, \eta}\hat{A}\ket{\psi, \eta}$ where the    wave function  for  each mode, 
 is that corresponding to the Bunch-Davies vacuum, which is just the harmonic oscillator ground state, at the start of inflation ($\eta=-\tau$) one has:
\begin{equation}
\begin{split}
    &\braket{p|\psi, -\tau}=\frac{1}{(\pi k)^{1/4}}e^{-p^{2}/2k},\\
    &\braket{x|\psi,-\tau}=\left(\frac{\pi}{k}\right)^{1/4}e^{-x^{2}k/2}.
\end{split}    
\end{equation}
 Note  that  we have used  lower case  $ ( x, p) $  to denote eigenvalues and  corresponding  eigenvectors  associated with the operators $\hat{X}$ a $ \hat{P}$.

For the field operator $\hat{X}$ and the field momentum operator, $ \hat{P}$, the equations of motion are:
\begin{equation}
    \begin{split}
        &\frac{d}{d\eta}\braket{\hat{X}}=\braket{\hat{P}}-\frac{\braket{\hat{X}}}{\eta}, \hspace{2cm} \frac{d}{d\eta}\braket{\hat{P}}=-k^{2}\braket{\hat{X}}+\frac{\braket{\hat{P}}}{\eta}\\
        &\frac{d^2}{d\eta^{2}}\braket{\hat{X}}=-\left(k^{2}-\frac{2}{\eta^{2}}\right)\braket{\hat{X}}, \hspace{2cm} \frac{d^2}{d\eta^{2}}\braket{\hat{P}}=-k^{2}\braket{\hat{P}},
    \end{split}
\end{equation}

and their solutions:
\begin{equation}
  \begin{split}
        &\braket{\hat{X}}=C_{1}\frac{-i}{k}e^{ik\eta}\left[1+\frac{i}{k\eta}\right]+C_{2}\frac{i}{k}e^{-i k\eta}\left[1+\frac{i}{k\eta} \right],\\
     &\braket{\hat{P}}=C_{1}e^{ik\eta}+C_{2}e^{-ik\eta}.
  \end{split}
\end{equation}
For the equations of the second degree, the following change of variables is made:
\begin{equation}\label{especificaciones}
      Q \equiv \braket{\hat{X}^{2}}, \hspace{1.2cm} R  \equiv \braket{\hat{P}^{2}}, \hspace{1.2cm} S \equiv \braket{[\hat{XP}+\hat{PX}]},
\end{equation}
and it follows that,
\begin{equation}\label{valexoec}
    \frac{d}{d\eta}Q=S-\frac{2Q}{\eta}, \hspace{1.2cm} \frac{d}{d\eta}R=-k^{2}S+\frac{2R}{\eta}, \hspace{1.2cm} \frac{d}{d\eta}S=2[R-k^{2}Q],
\end{equation}
whose solutions are
\begin{equation}\label{qsincolapso}
\begin{split}
    Q=&-C_{1}\frac{1}{k^{2}}e^{2ik\eta}\left(1+\frac{i}{k\eta} \right)^{2}-C_{2}\frac{1}{k^{2}}e^{-2ik\eta}\left(1-\frac{i}{k\eta} \right)^{2}+C_{3}\frac{1}{k^{2}}\left(1+\frac{1}{(k\eta)^{2}} \right),
    \end{split}
\end{equation}

\begin{equation}
R=C_{1}e^{2ik\eta}+C_{2} e^{-2ik\eta}+C_{3}   ,
\end{equation}

\begin{equation}
    S=-2iC_{1}\frac{1}{k}e^{2ik\eta}\left(1+\frac{i}{k\eta} \right)+2iC_{2}\frac{1}{k}e^{-2ik\eta}\left(1-\frac{i}{k\eta} \right)+C_{3}\frac{2}{k^{2}\eta},
\end{equation}
the boundary conditions, i.e, $\eta=-\tau$, 
\begin{equation}
    \begin{split}
        &Q(\mathcal{-\tau})=1/2k,\\
        &R(\mathcal{-\tau})=k/2,\\
        &S(\mathcal{-\tau})=0,
    \end{split}
\end{equation}
are used to evaluate the coefficients $C_{1}, C_{2}, C_{3}$:
\begin{equation}
    C_{1}=\frac{e^{2 i k \tau } (2 i k \tau -1 )}{8 k \tau ^2}, \hspace{0.5cm} C_{2}=C_{1}^{*} \hspace{0.5cm} C_{3}=\frac{1+k^{2}\tau^{2}}{2k\tau^{2}},
\end{equation}
when $^{*}$ denotes the complex conjugate. If one replaces the last in expression (\ref{qsincolapso}) one obtains, 
\begin{equation}
\begin{split}
 Q=&\frac{1}{4 \eta ^2 k^5 \tau ^2} \left[ \left(\eta ^2 k^2+1\right) \left(2 k^2 \tau ^2+1\right)+2 k \left(-\eta +\eta ^2 k^2 \tau -\tau \right) \sin (2 k (\eta +\tau ))\right. \\
 & \left. +\left(\eta  k^2 (\eta +4 \tau )-1\right) \cos (2 k (\eta +\tau ))\right],
\end{split}
\end{equation}
the last result, will result useful in the following section. 
\subsubsection{Quantum collapse scenario}
Now,  as in \cite{pearle}  we consider an adaptation to the context of  the   continuous spontaneous localization theory, CSL, which takes  as  the collapse operator $\hat{A}$ with a collapse rate $\lambda$. The dynamic of this theory it is given by two equations, the first is a modified of Shr\"odinger equation, whose solution is
 \begin{equation}
  \ket{\psi, t}= \mathcal{T}\exp{\left(-\int_{0}^{t}dt' \left[i\hat{H}+\frac{1}{4\lambda}[w(t')-2\lambda \hat{A}]^{2}\right] \right)}    \ket{\psi,0},
  \end{equation}
with  $\mathcal{T}$ the time-ordering operator, $w(t)$ a random classical function of time and the second equation is its the  probability of realization of a  stochastic function in the band   specified  by the  central values  $w(t_i)$ and  widths $dw(t_i)$\footnote{For more details   see the exposition  in \cite{ghirardi}.},   and  is given by 
    \begin{equation}
      PDw(t)\equiv \bra{\psi, t}\ket{\psi, t}\Pi_{t_{i}=0}^{t}\frac{dw(t_{i})}{\sqrt{2 \pi \lambda/dt}}.
  \end{equation}
%In certain instances, collapse can still occur. Alternatively, a state of equilibrium may be achieved, where the two dynamics reach a balance. In other cases, the unitary and non-unitary dynamics interact with each other in intriguing ways.

It is useful to possess a concise formulation for the density matrix that accurately represents the collective evolution within the ensemble. The evolution equation for the density matrix is given by the Lindblad equation:
\begin{equation}\label{lindbland}
    \frac{d}{dt}\rho(t)=-i[\hat{H},\rho(t)]-\frac{\lambda}{2}[\hat{A},[\hat{A},\rho(t)]].
\end{equation}
Consequently, the ensemble expectation value of the operator $\overline{\braket{\hat{O}}}=Tr \hat{O}\rho(t)$ satisfies the following expression:
\begin{equation}\label{nueva}
    \frac{d}{dt}\overline{\braket{\hat{O}}}=-i\overline{[\hat{O},\hat{H}]}-\frac{\lambda}{2}\overline{[\hat{A},[\hat{A},\hat{O}]]}.
\end{equation}

As mentioned before,  the modification  introduced by CSL dynamics tends  to collapse state vector toward eigenstates of $\hat{A}$ and it gives an ensemble of different evolutions of the state vector, each characterized by a particular realization of the   stochastic  function $w(t)$. In \cite{pearle} CSL is to be applied to the modes described by the focus operators $\hat{P}$, $\hat{X}$, and the Hamiltonian, $H_{k}=\frac{1}{2}\left[\hat{P}^{2}-\frac{1}{\eta}[\hat{P}\hat{X}+\hat{X}\hat{P}]+k^{2}\hat{X}^{2} \right]$. According with the ec. $(66)$ of the \cite{pearle} one obtains the following expression:
\begin{equation}
    \overline{\braket{\pi(k)}\braket{\pi(k')}^{*}}= \overline{\braket{P}^{2}}\delta{(\vec{k}-\vec{k'})},
\end{equation}
 The resulting  evolution  from the  initial state has the form $\braket{p|\psi,\eta}=e^{[-A(\eta)p^{2}+B(\eta)p+C(\eta)]}$, and  is  used  in  \cite{pearle}  with the aim is to find the spectrum, for that reason  $\overline{\braket{P}^{2}}$ is sought.  They compute $\overline{\braket{P^2}}$ and in turn the coefficients $A, B$ and $C$. Finally, using the probabilistic rule and the initial conditions it is found that the  ensemble  average is :
\begin{equation}\label{ppp}
  \overline{\braket{P}^{2}}=  \overline{\braket{P^2}} -\frac{1}{2(A+A^{*})}.
\end{equation}
From this result and using expression (\ref{nueva}),  the equations (\ref{valexoec}) are examined in two cases: when the field  acts  (namely $\hat X$)  as collapse operator  and when the conjugate momentum of the field   (namely $\hat X$)  plays that role . In this way, two results are found for the functional form of $k$ that  would  generate the required  scale invariant power   spectrum: for the first case  $\lambda=\tilde{\lambda}k$ and $\lambda=\tilde{\lambda}/k$ for the second  case. Afterwards, using these results, the authors investigate the overall scale  power spectrum and  find the typical  parameters  that  reproduce the observations from the CMB.

\subsubsection{Auxiliary calculations}

Since in this work we will focus in the case in which the quantum filed  itself acts as  collapse operator , it is useful to compute the Fourier transformation of the  expression for the  wave  function (of the generic mode under consideration).
%{to get $\braket{x| \psi,\eta}$ and reproduce equations [65-70] of \cite{pearle}}
\begin{equation}
  \begin{split}
        \braket{x|\psi,\eta}=&N \int dp \braket{x|p}\braket{p|\psi,\eta} \\
        &=N\int \exp(ipx)\exp[-A(\eta)p^{2}+B(\eta)p+C(\eta)]\\
        &=N\int \exp[-A(\eta)p^{2}+(B(\eta)+ix)p+C(\eta)].
  \end{split}
\end{equation}
%With the objective of simplifying the integral, one seeks a quadratic form, thereby obtaining:
%\begin{equation}
%\begin{split}
%        \psi(x,\eta)&=N\int dp \exp[\alpha(p+\beta)^{2}+\gamma]\\
 %       &=N \exp{\gamma}\int dp \exp[-\alpha(p+\beta)^{2}],
%\end{split}
%\end{equation}
%with $p'=p+\beta$ we get
%\begin{equation}
%     \psi(x,\eta)=N\exp{\gamma}\int dp' \exp[-\alpha p'^{2}],
%$\end{equation}
After   rather direct   calculation one finds 
\begin{equation}
   %\begin{split}
        \psi(x,\eta) % &=\frac{N}{\sqrt{A}}\exp\left[C+\frac{(B+ix)^{2}}{4A}
        %%\right],\\
        %&
        =\frac{N}{\sqrt{A}}\exp\left[\frac{-x^2}{4A}+\frac{ixB}{2A}+\frac{B^2}{4A}+C\right],
  % \end{split}
\end{equation}
which  has the    same functional  form as the original expression  $\braket{p|\psi, \eta}$,  so  one can write
\begin{equation}
 \braket{x|\psi, \eta}=N\exp{[-A'(\eta)x^{2}+B'(\eta)x+C'(\eta)]},
\end{equation}

and  identify the corresponding  coefficients:
\begin{equation}
    A'=\frac{1}{4A}, \hspace{1cm} B'=\frac{iB}{2A}, \hspace{1cm} C'=\frac{B^{2}}{4A}+C.
\end{equation}

%{which is obtain considering that momentum, $\hat{P}$, is the collapse operator  i.e, $\hat{A}=\hat{P}$ in (\ref{lindbland}). In this case the equation is:}
%\begin{equation}\label{puxx}
% \overline{\braket{X}^{2}}=   \overline{\braket{X^{2}}} -\frac{1}{2(A'+A^{'*})}.
%\end{equation}

%We proceed with a similar method as in \cite{pearle}, using the modified Schrödinger equation, the Fourier transformation to $\braket{p|\psi, \eta}$, i.e, 

%and in this way the corresponding expression has the same form as $\braket{p|\psi,\eta}$. 
%Since the goal is evaluate the ensemble average, $\overline{\braket{P}^2}$, we proceed to evaluate
Thus, one can use  the calculations in \cite{pearle} to obtain the analogous to expression for  $\overline{\braket{P}^2}=\overline{\braket{P^2}}-\frac{1}{2(A+A^{*})}$ and in this case, correspond to:
\begin{equation}
  \overline{\braket{X}^2}=\overline{\braket{X^2}}-\frac{1}{2(A'+A'^{*})}, 
\end{equation}
that in terms of $A$ it turns  into:
\begin{equation}
  \overline{\braket{X}^2}=\overline{\braket{X^2}}-\frac{2A A^{*}}{(A+A^{*})}. 
\end{equation}
For convenience we will use the first expression, i.e., in $A'$ terms. Using the modified Schr\"odinger equation it is possible obtain the coefficients $A'(\eta), B'(\eta), C'(\eta)$ and then calculate the expectation values of (\ref{valexoec}), keep in mind the previous change of variables (\ref{especificaciones}), such expectation values could be obtained from (\ref{nueva}). For the case in which $\hat{A}=\hat{X}$ it is obtain we get the coefficient $A$:
\begin{equation}\label{Aprima}
    \begin{split}
    A'(\eta)=\frac{\beta^2 \eta}{2i}\left[\frac{e^{2i\beta \eta}+C}{e^{2i\beta \eta}(1-i\beta \eta)+C(1+i\beta \eta)}\right],
    \end{split}
\end{equation}
being $\beta=\sqrt{k^2 -2i\lambda}$, and $C$ is determined by the initial condition $A'(-\tau)=k/2$
\begin{equation}
  C=-e^{2i\beta \tau} \frac{\beta^2 \tau-k\beta \tau+ik }{\beta^2 \tau+k\beta \tau+ik}.
\end{equation}
With the above mentioned it is possible to write the expectations values under CSL theory. 
One  complication  we face is that   while  in the original  work \cite{pearle} interest centered  on the behavior of the relevant  quantities only at  the end  of inflation  ,  for our purposes  we  need to the  behavior relevant quantities   throughout the inflationary regime.    The second order equations (\ref{valexoec}) are unchanged except for the $R$,
\begin{equation}\label{expectCSL}
\begin{split}
    &\dot{Q}=S-\frac{2Q}{\eta},\\
    &\dot{R}=-k^{2}S+\frac{2R}{\eta}+\lambda,\\
    &\dot{S}=2[R-k^{2}Q],
\end{split}    
\end{equation}
and the general solution is therefore the sum of previous homogeneous equations added an inhomogeneous solution part:
\begin{equation}
    Q=\frac{\lambda \eta}{2k^{2}}, \hspace{40pt} R=\frac{\lambda \eta}{2}, \hspace{40pt}  S=\frac{3\lambda}{2k^{2}}.
\end{equation}
The corresponding system is 
\begin{equation}\label{qconcolapso}
\begin{split}
    Q=&-\tilde{C_{1}}\frac{1}{k^{2}}e^{2ik\eta}\left(1+\frac{i}{k\eta} \right)^{2}-\tilde{C_{2}}\frac{1}{k^{2}}e^{-2ik\eta}\left(1-\frac{i}{k\eta} \right)^{2}+\tilde{C_{3}}\frac{1}{k^{2}}\left(1+\frac{1}{(k\eta)^{2}} \right)+\frac{\lambda \eta}{2k^{2}},
    \end{split}
\end{equation}
\begin{equation}\label{rconcola}
R=\tilde{C_{1}}e^{2ik\eta}+\tilde{C_{2}} e^{-2ik\eta}+\tilde{C_{3}}+\frac{\lambda \eta}{2}   ,
\end{equation}

\begin{equation}\label{sconcola}
    S=-2i\tilde{C_{1}}\frac{1}{k}e^{2ik\eta}\left(1+\frac{i}{k\eta} \right)+2i\tilde{C_{2}}\frac{1}{k}e^{-2ik\eta}\left(1-\frac{i}{k\eta} \right)+\tilde{C_{3}}\frac{2}{k^{2}\eta}+\frac{3\lambda}{2k^{2}}.
\end{equation}
%which is canceled for $\lambda=0$, where $\mathcal{T}$ is the time according to the beginning of the inflation season. For the spectrum to be scale invariant, it is necessary that the dominant term goes as
%\begin{equation}\label{propK}
%    \lambda=\Tilde{\lambda}/k,
%\end{equation}
In the same way that we proceed in the collapse-free case we use the initial condition to find the coefficients $\tilde{C_{1}}$, $\tilde{C_{2}}$ and $\tilde{C_{3}}$:
\begin{equation}
\begin{split}
    &\tilde{C_{1}}=\frac{e^{2 i k \tau } \left(2 i k^2 \tau -i \lambda  k \tau ^2+2 \lambda  \tau -k\right)}{8 k^2 \tau ^2}, \\
    &\tilde{C_{2}}=\tilde{C_{1}}^{*}, \\
    &\tilde{C_{3}}=\frac{2 k^2 \tau ^2 (\lambda  \tau +k)-2 \lambda  \tau +k}{4 k^2 \tau ^2},
    \end{split}
\end{equation}
now substituting  
the above expressions one  finds the following spectrum:
\begin{equation}\label{domlamt}
      \overline{\braket{\hat{P}}^2}=\frac{\lambda \mathcal{T}}{2}-\frac{\sqrt{k^4+4\lambda}}{\sqrt{k^2-2i\lambda}+\sqrt{k^2+2i\lambda}},
\end{equation}
for the case $\lambda=0$ the expression vanishes. The requirement for this spectrum to be scale invariant is that the dominant term goes as
\begin{equation}\label{colcampo}
    \lambda=\Tilde{\lambda}k,
\end{equation}
with $\tilde{\lambda}\approx 10^{-5}Mpc^{-1}$.

\subsubsection{Useful estimations}
We proceed to review the magnitude estimates that will justify the approximations that will be made in this work. We start by saying that temperature fluctuations go like
\begin{equation}
    \frac{\Delta T}{T}=\frac{1}{3}\psi \approx 10^{-5},
\end{equation}
Considering the dominant term, we have
\begin{equation}
    \left[\frac{\Delta T}{T} \right]^2 =\frac{\pi}{4}\left[\frac{4\pi G \phi_{0}'}{3a} \right]^2 \tilde{\lambda}\tau\mathcal{I},
\end{equation}
where $\mathcal{I}$ represents a modes sum $k$, $\mathcal{I}= \int_{10^{-3}}^{10^2} dk/k\approx 11.5$, the integration limits represent the observed range for $k$, $10^{-3}Mpc^{-1}<k<10^{2}Mpc^{-1}$.\\ 
As discussed in \cite{hanno}, the effect on the reheating period is given by:
\begin{equation}
   \left( \frac{\Delta T}{T} \right)^{2} \approx \frac{1}{\epsilon}\frac{V}{M_{p}^4}\tilde{\lambda}\tau\mathcal{I}.
\end{equation}
From the observations we have $\left(\frac{\Delta T}{T}\right)^2\sim 10^{- 10}$. The deviation from the flatness of the spectrum tells us that $\epsilon\approx 10^{- 2}$. With $V^{1/4}$ thought by the scale of GUT $\approx 10^{15}GeV\approx 10^{- 4}M_{p}$ and taking $\mathcal{I}\approx 10$. It can be concluded that $ \tilde {\lambda} \tau $ is of the order of $10^{3}\gg1$. \\
T spectral index $n_{s}$ usually  involves the first  two slow-roll parameter, however s  for  the case  of  the chaotic potential $V=\frac{1}{2}m^2 \phi^2$ they    coincide  \footnote{Considering  \cite{Copeland:1997mn} that $n-1=-6\epsilon+2\eta$ with   the  expressions for   first  two slow roll parameters given by  $\epsilon=\frac{M^2_{P}}{16\pi}{\left(\frac{V'}{V}\right)^2}$ and $\eta=\frac{M^2_{P}}{16\pi}{\left(\frac{V''}{V}\right)}$,   one  readily   finds their equality for the  case at  hand.}.

Now that the modified  account  of   the emergence of structure has been  presented, we move on to the study of a problem that of eternal inflation. We will recall  the ``standard '' version of this problem, and we  then will  offer  our  critique and  modified  analysis.
%%%%%%%%%% AQUÍ TERMINAAAAAAAAAA %%%%%%%%

\section{Eternal Inflation} 
Let  us   recall  the eternal inflation problem is as it was presented for instance  in \cite{linde}.
%According to \cite{linde} is
The   considerations  start  with   the zero mode of inflation field $\phi_{0}$   which is rolling  slowly towards the potential's minimum,  while   is also       subject to ``quantum  fluctuations"  which are  considered  as  stochastically in nature.  These two effects  are  compared  by   considering the  classical  displacement over  a characteristic time, $t_{char}$, in which the field undergoes a classical shift $\Delta \phi_{0}$ with the magnitude of  to quantum  uncertainties or fluctuations, $\delta_{quan}\phi_{0}$. 
The  two possibilities arise:
     \begin{equation}\label{sinp}
           |\delta_{quan} \phi_{0}| \leq \Delta \phi_{0}
     \end{equation}
     \begin{equation}\label{updown}
           |\delta_{quan} \phi_{0}| \geq \Delta \phi_{0}
     \end{equation}
      In the first case %From (\ref{sinp})
      one  concludes that despite the fluctuations the  filed   will effectively  move downwards  towards the potential minimum, where   inflation would end.  However  in the second case % (\ref{updown}) we notice that 
      the   stochastisity of the dominant term would imply that in some places the field will move  up  away  from the   minimum of the potential   while  and in others it will move  downwards.
      The point is that in those regions where the filed moves upwards, the rate of cosmological expansion increases, which it decreases and in the regions  where it goes downwards.
      Thus the regions where the field went up will grow more  than the and the regions where it went downwards  and thus, 
     over time the universe will be predominantly covered by regions where there was more expansion.
   Thus, the inflationary potential grows and inflation never ends.
   
   As  we  argued at the start of  this  work,   there are flaws  in this  story that, however reappear now in   a fully  justified fashion as  a result of the  incorporation of a truly stochastic   aspect to the dynamics of the quantum field,     which  we  did  in order to account for the emergence of cosmic seeds. That is, the problem  re-emerges   as  a result of the  inclusion of   spontaneous collapse theories.   

\section{A strategy to address the problem}

The inclusion into the inflationary  account  of  the behaviour of the early universe of a  spontaneous collapse theory  brings  in   both,  an  explanation for the breaking of homogeneity and isotropy of the vacuum state of the inflaton  field,  a  justification for the  stochasticity which   otherwise   seems unfounded, and  the potential to  resurrect the   problem of  eternal inflation. However it also brings  into the  dynamics  new elements  which might  be adjusted   so  as to deal  with the  last problem. 
 We  will thus  consider  a  particular treatment of that kind  together  with  the lessons  learned regarding  the  empirical adequacy of the approach regarding the   predictions  for the   power  spectrum of  scalar perturbations reflected both in the CMB  and  the  studies of BAO \cite{KSPLANK, BAO?},  and  study  possible  ways to make   all that compatible    with having no   eternal inflation. 
   
    As we mentioned before in order to  specify  the version of the adaptation of   that  kind of  theories  to the   situation  involving  relativistic quantum fields and   gravitation,  one must  among other  specify the quantity that    one assumes  plays the  of  the collapse operator \footnote{One  would need of course to   do much  more  when  looking for a truly fundamental theory,  including a  formulation that is      fully  generally  covariant  and where the collapse operator  u operators  are  chose n once  and far all.  although    important progress  has been made in recent years (for instance,   with the formulation of   relativistic  versions of  collapse theories (\cite{Bedingham1, Bedingham2, Tumulkagrw},  and the development of  formal  schemes to put together  collapse theories  with semi-classical gravity\cite{Juarez-Aubry:2019jon, Juarez-Aubry:2021abq})  it is   fair to say that the task is  unfinished. Nonetheless  we  already can show, as     with the present   analysis the potential  such   theories have  to deal  with various   open problems  facing some  areas of  theoretical physics \cite{Bengochea2020efe, PROBLEMBHINFORMATION, EVAPORATION}.}.  In  the work \cite{pearle} two proposals are  studied :  one  in   which the collapse operator role  is taken  by  momentum conjugate  to  the field,  and a second  in  which it is  the  quantum filed  itself that plays  that role.  In the  analysis  it is found  that    in order  to  generate   a primordial spectrum  of  stochastic  density fluctuations that  is  essentially  scale invariant (as  required for  basic  adjustment to empirical  data)  the collapse rate  must depend on the  wave number  $k$ of the mode. In fact what is  required   in  each case is  $\lambda\sim\tilde{\lambda}/k$ and $\lambda\sim\tilde{\lambda}k$, respectively.
   
 The problem of eternal inflation using collapse theories and semi-classical gravity has been address before in \cite{Leon:2017sru}, considering the case  momentum as the collapse operator of the theory. %and it is shown a solution to solve the problem.  
 Here we  will  focus  on the  case   where the quantum filed  filed plays the role  of collapse operator.
 We  find that  proposal   more  attractive  
 % with the zero mode when momentum collapses the present work is based on the result in that field is which collapses.
because  taking   $\lambda=\tilde{\lambda}k$ the collapse  rate  for zero mode  vanishes,   thus  eliminating from the onset  the  collapses that could  generate  the eternal inflation problem. %However a  careful  considerations indicate the  analysis
%needs to more complex because  consideration must be given to  the modes whose wave-length is  so  long that, or  are near to the zero mode,  so  as to effectively act like  it in practice These are  identified as  modes that have a physical wave-length $\lambda_{phys}$ greater than a critical value $\lambda_{C}$. For this critical variables we will considered two natural alternatives: the one determined by the particle horizon, and the one corresponding to the inverse Hubble radius. 
 However one  further consideration one  notes   that  modes that  have  wavelengths that  are too large to be  associated  with   causal  effects during inflation,    would  have to be considered  as  representing  {\it  ``effectively  zero  modes''}.  That is,  those  would be  the   modes with  wavelengths  that are larger than the particle horizon. In  most   treatments  of the issue of   eternal  inflation,  one focuses on what takes place  {\it withing individual  Hubble volumes} , and  thus,  the magnitude   of the quantum fluctuation  that  is  compared  with the classical displacement  is    estimated  by  focusing on  the modes with wavelength that is larger than the Hubble radius (leading the ``quantum fluctuation'' as $\braket{\phi^2}^{1/2} \sim \frac{H}{2\pi}$  (see \cite{Brand} for  an explicit calculation of a similar  form). We do not consider that  this  would be  the appropriate  criterion to be employ in our analysis, 
 as we have emphasized that  the  quantity  driving inflation  is  the   zero mode of the inflaton field,  which   is by definition homogeneous. Therefore   any   type of fluctuation  (or indeterminacy) pertaining to it,   would represent  something affecting the   whole universe,  and not just one Hubble  volume  within it.    However on  further consideration one  notes   that  modes that  have  wavelengths that  are too large to be  associated  with   causal  effects during inflation,    would  have to be considered  as  representing  {\it  ``effectively  zero  modes''}.  That is,  it is  modes with  wavelengths   that are larger than the particle horizon, (which is  what  determines  causal  contact), that  must be   considered as  {\it ``effectively  zero modes''}. 
 The Hubble radius  is a good  estimate of regions in  which a cosmic  fluid  could be  expected to be in thermal equilibrium, but not  necessarily  as  characterizing  the  regions   over which  quantum  entanglement  and correlations exists.   The   particle horizon  is  what  determines  causal  contact, thus  determines  which   modes with must be   considered as  {\it ``effectively  zero modes''}. 
We  thus  analyze  their   effect and  
ascertain  under what conditions    they  lead  the eternal inflation problem  and  how  is  it possible to  evade it. \footnote{It  is  worthwhile  emphasizing that  throughout this   analysis  we are   working under the assumption that,  expect for the zero mode,  all modes  of the  inflaton  field can  be  considered    to be  initially found in  the  adiabatic   vacuum, an assumption that might be taken  as  natural  in the  present context,   but for  which there  does not  seem  to exist  anything close to  a  strict justification.}.
%  The general lesson here 

%ACA.

 %The Newtonian potential for the relevant modes is constant as function of $x$, thence is possible write the perturbed FLRW metric as: 
% \begin{equation}
 %   \begin{split}
 %   dS^{2} &= a^2 (\eta)   [ - (1 +  2 \psi ( \eta) )  {d  \eta } ^2  +  (1-  2\psi(\eta)) \delta_{ij} d x^i dx^j ] 
%   \\ &= a^2 (\eta)(1-  2\psi(\eta))   [  - (1 +  2 \psi ( \eta) )(1-  2\psi(\eta))^{-1}  {d  \eta }^2  +   \delta_{ij} d x^i dx^j ] 
  % \\ & =   \tilde a^2 (\tilde \eta) [ - d {\tilde  \eta}^2 + \delta_{ij} d x^i dx^j ],
   %  \end{split}
%     \end{equation}
%the temporal variable becomes to $ d {\tilde  \eta} =[(1 +  2 \psi ( \eta) )(1-  2\psi(\eta))^{-1}]^{1/2}  {d  \eta } $  and the scale factor is given by $\tilde a = a (1- 2\psi )^{1/2} $. 

%In consequence, one could notice that the  modes  whose $k$ is small enough could have an impact as the zero mode, it means the behaviour of scale factor could influence in the causal connected region, i.e, on the particle horizon region, $D_{HP}$. In the reference \cite{waldo} it is shown as the correlations could beyond the horizon but in this work we do not focus on it.  \\
%\subsection{$\lambda_{C}$ corresponding to the particle horizon}
The size  of  particle horizon in  the context of interest is:  
\begin{equation}
D_{HP}=a(t)\int_{t_{i}}^{t} \frac{1}{a(t)}dt,    
\end{equation}
where $a(t)=C\exp{H_{I}t}$.  Thus:
\begin{equation}
\begin{split}
    D_{HP}&=\frac{1}{H_{I}}\left[ \exp{H_{I}(t-t_{i})}-1\right]\\
    &\approx \frac{\exp{H_{I}(t-t_{i})}}{H_{I}}.
\end{split}
\end{equation}
The modes that could give rise to  the eternal  inflation  problem  are then those   for which :
\begin{equation}
    \lambda_{phys}=a(\eta)\frac{2\pi}{k}>>\frac{\exp{H_{I}(t-t_{i})}}{H_{I}}.
\end{equation}
This  implies that, 
\begin{equation}
    \begin{split}
        \frac{2\pi}{k}&>> \frac{1}{H_{I}a(t)}\left[\frac{C\exp{H_{I}t}}{C\exp{H_{I}t_{i}}}\right]\\
        &=\frac{1}{H_{I}a(t)}\left[\frac{a(t)}{C\exp{H_{I}t_{i}}}\right]\\
        &=\frac{1}{H_{I}a(t_{i})},
    \end{split}
\end{equation}
taking as  initial condition, $a(\tau)=\frac{1}{H_{I}\tau}$, we have
\begin{equation}
   k<<k_{c}=\frac{2\pi}{\tau}.
\end{equation}
The next step is estimate the  magnitude of the   cumulative  stochastic  displacement of the expected value of the field associated with the collapse of these modes, $\overline{{\langle  X \rangle }^2} =
  \overline{{\langle  X^2 \rangle }}-\frac{1}{2 | Re A'(\eta)|}$.  The first part, $  \overline{{\langle  X^2 \rangle }}$,  is obtained from (\ref{qconcolapso}) :
\begin{equation}
\begin{split}
 \overline{{\langle  X^2 \rangle }}=\frac{1}{8 \eta ^2 k^6 \tau ^2}& \left[ 4 \eta ^3 \lambda  k^4 \tau ^2+(\eta  k-i)^2 e^{-2 i k (\eta +\tau )} \left(2 i k^2 \tau -i \lambda  k \tau ^2-2 \lambda  \tau +k\right) \right. \\
 &\left. +(\eta  k+i)^2 e^{2 i k (\eta +\tau )} \left(-2 i k^2 \tau +i \lambda  k \tau ^2-2 \lambda  \tau +k\right)\right. \\
 & \left. + 2 \left(\eta ^2 k^2+1\right) \left(2 k^2 \tau ^2 (\lambda  \tau +k)-2 \lambda  \tau +k\right) \right].
 \end{split}
\end{equation}  
In obtaining the above result,   eqs (\ref{rconcola})-(\ref{sconcola}) of \cite{pearle} were used. Next we evaluate  the second term making  use the expression:
\begin{equation}\label{Aprima}
    \begin{split}
    A'(\eta)=\frac{\beta^2 \eta}{2i}\left[\frac{e^{2i\beta \eta}+C}{e^{2i\beta \eta}(1-i\beta \eta)+C(1+i\beta \eta)}\right],
    \end{split}
\end{equation}
with $\beta\equiv\sqrt{k^2 -2i\lambda}$, 
\begin{equation}
    C=-e^{2i\beta \tau} \frac{\beta^2 \tau-k\beta \tau+ik }{\beta^2 \tau+k\beta \tau+ik}.
\end{equation}
From the   equations above we find the following expression for $ Re [A '(\eta)] $:
\begin{equation}\label{de}
    \begin{split}
    & Re[A(\eta)]=\frac{\eta}{8u}\exp{\left(-\frac{2kT}{u}-2i\tau u\right)}\\
&\quad \quad \quad \Bigg[ \exp{(-2i\eta u)} \Big( \tau (k-iu^2)(-iku+k-iu^2) + iku^2 \Big)  \nonumber \\
& \quad \quad \times \Big( k \exp{\left(\frac{2kT}{u}\right)} \Big( \eta(k^2+u^4)-2iu^3 \Big) \Big( ku^2+\tau(u^2-ki)(u^2+k(u+i)) \Big) \nonumber \\
&\quad \quad  -u^2\exp{\left(\frac{4kT}{u}+2iT u\right)} \Big( \eta(k^2+u^4)-2ku \Big) \Big( \tau(u^2-ik)(u^2+k(u-i)) + iku^2 \Big) \Big) \nonumber \\
& \quad \quad + \exp{(2i\tau u)} \Big( \tau (u+ik)(u^2+k(u-i)) + iku^2 \Big) \Big( k \exp{\left(\frac{2kT}{u}+2iT u\right)}\\
&\quad \quad \times \Big( \eta(k^2+u^4)+2iu^3 \Big) \nonumber 
   \Big( -ku^2+\tau(k+iu^2)(u^2+k(u-i)) \Big)\\
&\quad \quad  + iu^2 \Big( \eta(k^2+u^4)+2ku \Big) \Big( ku^2+\tau(k+iu^2)(u^2-ki) \Big) \Big) \Bigg] \nonumber \\
& \quad \quad \times\Big[8u \Big( u\sinh\Big(\frac{kT}{u}+iT u\Big) \Big(-ku^2-\tau(\eta k+i)(k+iu^2)^2\Big) \nonumber \\
& \quad \quad + (k+iu^2)\cosh\Big(\frac{kT}{u}+iT u\Big) \Big(\eta ku^2+\tau\big(ku^2+i\eta(k+iu^2)^2\big)\Big) \Big) \nonumber \\
& \quad \quad \times \Big( (k-iu^2)\cosh\Big(\frac{kT}{u}-iT u\Big) \Big(-i\eta k^2\tau+ku^2(T-2\eta\tau)+i\eta\tau u^4\Big) \nonumber \\
& \quad \quad +u\Big( \tau(1+i\eta k)(k-iu^2)^2 + iku^2 \Big) \sin\Big(\frac{T(u^2 + ik)}{u}\Big) \Big) \Bigg]^{-1}    
    \end{split}
\end{equation}
where, $T=\eta+\tau$ and $u=\sqrt{k^2+\sqrt{k^4+4 \lambda ^2}}/\sqrt{2}$. If $ \lambda=0$.  We obtain:
\begin{equation}
\begin{split}
    A'(\eta)^{-1}|_{\lambda=0}&=\frac{1}{\eta ^2 k^5 \tau ^2}\\
   & \times\left[\left(\eta ^2 k^2+1\right) \left(2 k^2 \tau ^2+1\right)+2 k \left(\eta  \left(\eta  k^2 \tau -1\right)-\tau \right) \sin (2 k (\eta +\tau ))\right.\\
    &\left. +\left(\eta  k^2 (\eta +4 \tau )-1\right) \cos (2 k (\eta +\tau )) \right],
    \end{split}
\end{equation}

\begin{equation}
\begin{split}
  \overline{{\langle  X^2 \rangle }}|_{\lambda=0}&= \frac{1}{4 \eta ^2 k^5 \tau ^2}\\
  &\times \left[\left(\eta ^2 k^2+1\right) \left(2 k^2 \tau ^2+1\right)+2 k \left(-\eta +\eta ^2 k^2 \tau -\tau \right) \sin (2 k (\eta +\tau ))\right.\\
   &+\left.\left(\eta  k^2 (\eta +4 \tau )-1\right) \cos (2 k (\eta +\tau ))\right].
    \end{split}
\end{equation}
Note  that  we have: $\overline{\langle X \rangle^2}_{\lambda=0}=0$, which is  as  expected. Namely  if there is no  collapse the  quantum expectation value of the harmonic oscillator  continues to vanishes at all times  if it (and  that of its momentum conjugate) vanished initially.
Now we  expand $\overline{\langle X \rangle^2}$ in Taylor series, up  to the first order  in $\lambda$ , and to the eight order  in $k$ and obtain
\begin{equation}\label{REFERR}
 \overline{\langle X \rangle^2} \approx  -\frac{\lambda \tau^{3}(\eta + \tau)}{\eta^{3} k^{2}}.
\end{equation}
%it is important remark that the last expresion must satisfies two conditions: 1) the inital condition, i.e, $\eta=-\tau$, 2) $\overline{\langle X \rangle^2}\geq0$ and $\overline{\langle X \rangle^2}=0$ for $\lambda=0$. From (\ref{REFERR}) it is clear that the first condition is valid because when $\eta=-\tau$ the expresion is vanished. To show the valid of 2) we can notice that $\eta$ is a negative quantity while $\tau$ is possitive, in fact, $-\tau<\eta<-\eta_{f}$, and in the estimates section of \cite{pearle} one can check that $\tau=10^{8}MpC$ and $\eta=10^{-22}MpC$. Then,  the quantity $-\frac{\tau^{3}}{\eta^{3}}>0$ and also $(\eta+\tau)>0$, with the above we have check that $\overline{\langle X \rangle^2}$ is possitive defined. Also, it is easy to see from (\ref{REFERR}) that $\lambda=0$ implies $\overline{\langle X \rangle^2}=0$.

Using that $\delta \phi = a^{- 1}y$ and  the expression  for the  Fourier mode of this field $\tilde{\hat{y}}_{\vec{k}}={\hat {X}}({\vec k})$ (recalling we have   employed the  variable $\hat{X}_{\vec{k}}=\sqrt {d^3k}{\hat{X}}({\vec k})$ where $ d ^ 3k $ represents an ``infinitesimal '' volume in the space of the $k$'s around $ \vec{k} $). We can now proceed to make the desired estimate.

However  before doing so,  and anticipating that  the  direct extrapolation of the  simple  for of the  $ k$  dependence   of  $\lambda$  encountered in  the previous  studies might not be   adequate  for the task at hand, we consider a  slight generalization In fact  as argued in previous  works  works  one might expect that  the  current versions    of the collapse  theory  are  just  effective  versions of  a more fundamental theory,  it  is  likely  that  what  we  take  as  the collapse  parameter   $\lambda$  might be  an   effective  characterization of   a quantity that depends on various other quantities including  space-time  curvature. This suggests that the collapse rate could exhibit variation depending on the specific context at hand.  In fact   it was  already noted  in  \cite{pearle}  in order to  avoid the divergences of the stochastic  fluctuations  of  field itself  something like, going beyond the  simple  dependence  already mentioned  that seems  to be required.  Thus  we   consider  of  form  of  collapse rate  that, adjust to  what is  needed  to    account  for the  essentially  flat   power  spectrum consistent  with  CMB and BAO   data  with some  added  flexibility at   very low $k$ so as  deal  with  the issue of eternal inflation. Since we  are considering that $\hat{X} $ acts as  the collapse operator  we  need  $\lambda \sim \tilde{\lambda}k$,  
so  we  take  a slightly more  general form:
\begin{equation}\label{newprop}
    \lambda=\tilde{\lambda}\left(\frac{k^{\alpha + 1}}{(b+k)^{\alpha}}\right),
\end{equation}
where $b$ has the same units of $k$.   Note that for  $\alpha=0$ we  recover  $\lambda=\tilde{\lambda}k$. The last expression also  reduces to $\lambda=\tilde{\lambda}k$ in the limit $k\gg b$, 
so we  will recover  an empirically adequate  power spectrum   provided $k\gg b$  for all  $k$ corresponding  to  the modes  relevant  for the CMB   and  BAO  data.   On the other hand  we  expect  that for the modes  that could play a role in  eternal inflation  $b \gg k$ so that for those we have, 
\begin{equation}
    \lambda=\tilde{\lambda}\left(\frac{k^{\alpha+1}}{b^\alpha}\right).
\end{equation}
We assume that the displacements in the ``discrete '' variables $ \hat{X}_{\vec{k}}$ are statistically independent (since the stochastic processes of CSL dynamics are assumed independent for each mode), so:
   \begin{equation}\label{u5}
 \overline {\langle \delta \phi \rangle^2} =   a(\eta)^{-2}  \overline {\langle y_{k} \rangle^2} =  a(\eta)^{-2}  \sum_{ k < k _{c}} \overline {\langle y_{k} \rangle^2} =  a(\eta)^{-2}  \int_{0}^{ k _{c}}d^3 k \overline {\langle X (k) \rangle^2}.
\end{equation}
Replacing  in last expression the result (\ref{REFERR}) we find :
\begin{equation}\label{psustituir}
 \overline {\langle \delta \phi \rangle^2} =   a(\eta)^{-2} \overline {\langle y_{k} \rangle^2} = a(\eta)^{-2}  \sum_{ k < k _{c}} \overline {\langle y_{k} \rangle^2} =  a(\eta)^{-2}  \int_{0}^{ k _{c}} 4\pi k^{2} dk  \left[   -\frac{\lambda \tau^{3}(\eta + \tau)}{\eta^3 k^{2}} \right].
\end{equation}
%%%%%%% DATOS DE ESTIMACIONES COMENTADOS%%%%%%%%%
%where $M_{p}\approx 10^{19} GeV$, $m=\phi_{0}\approx 10^{15}GeV$, $\tilde{\lambda}\tau \approx 10^3$, and
%$(m^2\phi_{0}^{2}(\tilde{\lambda}\tau)/M_{p}^{4})\approx 10^{-9}$, $\tau \approx 10^8 Mpc$. 
%Then, the requirement is $\eta>10^{33}MpC$ but, $\eta \in (10^{8}MpC, 10^{-22}MpC)$. It is clear that the problem does not solve. Nevertheless, as we mentioned in the last sections one notable benefit of that theories is the possibility to adjust the free parameters of the theory as the collapse rate. In the next section, we will present a proposal in which the collapse rate was modified in a strategic way.

 Evaluating the  integral  we find:
\begin{equation}\label{INCERQUAD}
\begin{split}
    \overline{\braket{\delta \phi}^2}&=-4\pi a(\eta)^{-2}\frac{\tau^3}{\eta^3}(\eta+\tau)\tilde{\lambda}\int_{0}^{2\pi/\tau}\frac{k^{\alpha+1}}{b^\alpha}\\
    &=-4\pi a(\eta)^{-2}\frac{\tilde{\lambda}\tau^3}{\eta^3}(\eta+\tau)\frac{1}{b^\alpha} \frac{1}{\alpha+2}k^{\alpha+2}\Big|_{0}^{k_{c}}.
\end{split}
\end{equation}
Using  (\ref{psustituir})  we  obtain, 
\begin{equation}\label{cota}. 
    \overline{\braket{\delta \phi}^2}
    =(4\pi)^{3}H_{I}^{2}(\tilde{\lambda}\tau)\left(\frac{\tau}{-\eta}-1 \right)\frac{1}{\alpha+2}\left(\frac{2\pi}{b\tau} \right)^{\alpha},
\end{equation}
%%%%}

where we used  $a(\eta)=-1/H_{I}\eta$. We define the   quantum  displacement  $\Delta Q=\sqrt{\overline{\braket{\delta \phi}^2}}$  an compared with the classical displacement over a Hubble time,  arising from slow-roll  regime (\ref{slowr-cond}),  
%\textcolor{blue}{$\Delta C\equiv\left(\frac{d}{d\eta}\phi_{0}\right)t_{Hub}$}, \textcolor{red}{
$\Delta C\equiv\left(\frac{d}{d\eta}\phi_{0}\right)\eta_{Hub}$ which can be   expressed as: 
\begin{equation}
 \Delta C=\frac{2m^2}{3H_{I}^{2}}\phi_{0}e,
\end{equation}
where %\textcolor{blue}{$t_{Hub}=\eta e$} \textcolor{red}{
$\eta_{Hub}=\eta e$ has been used\footnote{This is the
period over which the universe change its size by an factor of order $e$. It is the usual notation but in terms of the conformal time.}. Thus  the condition for avoiding eternal inflation is:
\begin{equation}
  \sqrt{\overline{\braket{\delta \phi}^2}}<\frac{2m^2}{3H_{I}^{2}}\phi_{0}e, 
\end{equation}
and  equivalently :
\begin{equation}
  {\overline{\braket{\delta \phi}^2}}<\frac{4m^4}{9H_{I}^{4}}\phi_{0}^2 e^2. 
\end{equation}
Using (\ref{INCERQUAD}) in the last expression and  rearranging  terms, the condition takes the form:
\begin{equation}
    (4\pi)^3 H_{I}^2 (\tilde{\lambda}\tau)\frac{(\tau+\eta)}{|\eta|}\frac{1}{\alpha +2}\left(\frac{2\pi}{b\tau} \right)^{\alpha} < \frac{4m^4}{9H_{I}^4} \phi_{0}^2 e^2
\end{equation}
Rewriting the previous expression and using that  $H_{I}=(8\pi m^2 \phi_{0}^2 / 2M_{p}^{2})^{1/2}$ and the estimates:  $M_{p}\approx 10^{19} GeV$, $m=\phi_{0}\approx 10^{15}GeV$,  so $\frac{m^2 \phi_{0}^{2}}{M_{p}^4} \approx 10^{-16}$, $\tilde{\lambda}\tau \approx 10^{3}$, $\frac{\tau}{\eta}\approx 10^{30}$ we obtain the condition:

\begin{equation} \label{condifin}
    10^{15}\frac{1}{\alpha +2}\left( \frac{2\pi}{b \tau} \right)^{\alpha}<1.
\end{equation}

 Note, that, as  anticipated the simple case   where $ \alpha = 0$ does not avoid the    eternal  inflation problem. In other words, the above is the condition that    avoids   the  eternal inflation problem. We must then look   for $\alpha$ and $b$ such that $b<k$ for $ k$ in the range, $10^{-3}Mpc^{-1}\leq k \leq 10^{2}Mpc^{-1}$ \cite{pearle}. 
There values of  $(\alpha, b)$ with $b \leq 10^{-3}Mpc^{-1}$. 
Here  we  present a plot  showing  an interesting  region in which the eternal  inflation   problem does not arise:

\begin{figure}[h!]
\includegraphics[scale=.60]{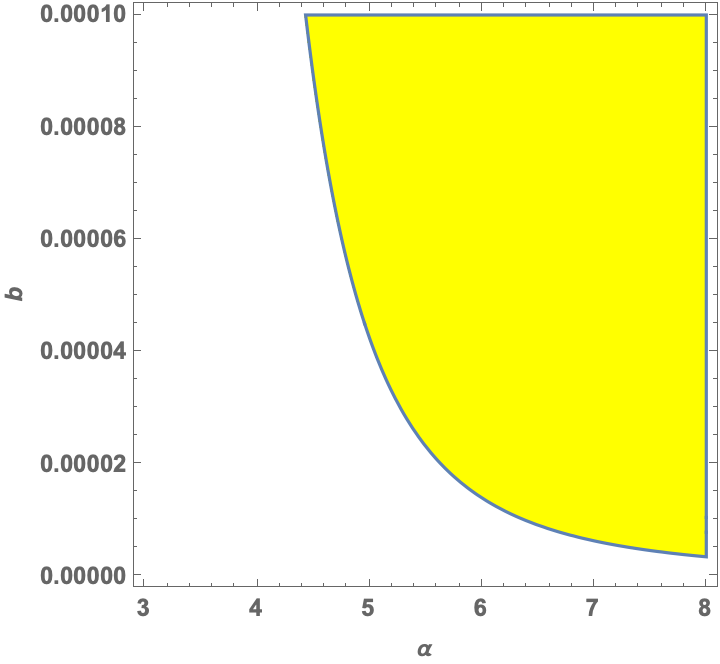}
\centering
\caption{In the last plot, the units are in $Mpc^{-1}$, the yellow region represents where there is NO eternal inflation problem. The $\alpha$ scale was chosen arbitrarily while $b$ was selected to clearly  comply  with  CMB  and BAO  bounds.}
\end{figure}

\newpage

\newpage
\newpage

\section{Conclusions} \label{s:numerical}
In this work, we have analyzed the aspects that could contribute to the re-emergence of the problem of eternal inflation in  the context in  which a  spontaneous   collapse theory has  been added to the  standard inflationary account of the emergence of the  primordial  seeds of cosmic  structure . We began by recognizing that within the framework of the standard theory, this problem is no present in the    usually  stated  form  once  one recognizes that  the zero mode of the field is independent of position. Furthermore, once  one notes that quantum fluctuations can  not justifiably be identified  with  statistical fluctuations, one  again  concludes that  the problem does not exist, however  that  will also   invalidate  the standard  account  for  the  origin of  cosmic. 
Based on a study that incorporates a new element accounting for the breaking of homogeneity and isotropy \cite{pearle}, we center our study around this aspect. In other words, if one seeks a consistent treatment to generate the primordial structure, the price to be paid is the reappearance of the problem of eternal inflation. Once again, focusing on  one  of the cases  studied in  \cite{pearle}, we note that the collapse rate is proportional to the wave number  of  mode of the field, and therefore, there would be no collapse for the zero mode of the field, seemingly   resolving the  problem   outright. However, there are other modes to consider, those  whose wavelength are close  enough to  to the zero mode.  We identified  the modes that have a physical wavelength greater than the particle horizon as  those that could have an effect on the eternal inflation problem. We examined the  effect  of   these modes utilizing the proposed collapse  theory  while  considering  a  slightly  modified rate that takes into account both the observable modes and those that could give rise to eternal inflation. At first glance, modifying this rate may appear unjustified. However, as previously mentioned, previous studies \cite{Bengochea2020efe, Modak:2014vya} provide clues that the collapse rate could be an effective  parameter  emergent from a more  fundamental theory and  effectively dependent on curvature, suggesting that it could vary from context  to context where the collapse theory is applied. We find  a   condition on the  parameters  characterizing the  the modified  collapse rate, for which the eternal inflation problem does not arise. It  is  worthwhile  emphasizing that  although the present analysis   concerns  a very specific   version of  a spontaneous collapse theory and  a  rather {\it add hoc} parametrization of  the collapse rate, the central point is  that the search for a   satisfactory  account for  emergence of the primordial seeds of cosmic  structure through the inclusion of   that type of modification  of   the standard  versions of quantum theory, which in turn, are motivated by the  quest  for a consistent
treatment of the  measurement problem, have the potential  to  offer unexpected  additional   advantages.
In this case it seemed  that  the  implicit    stochasticity  in the dynamics  of  theory would lead  back to the problem  of eternal inflation,  but  as  a result  of  the  flexibility
provided by the unknown exact form  of  the collapse theory, adjustments can be made to  remove the  problem. The  situation is  reminiscent of  similar  situations found  in  the  works \cite{Modak:2014vya, Bengochea2020efe, Rodriguez:2017rjh, Juarez-Aubry:2022qdp}. 
\section*{Acknowledgements}

R. L. Lechuga    Support from a   CONACYT (M\'exico)   doctoral  fellowship.
D. Sudarsky acknowledges partial financial support from  CONACYT (M\'exico) grant  No 140630. Additionally, we express our gratitude to the referee for their valuable observations.

\newpage

%\bibliographystyle{unsrt}
%\bibliography{eternal}

\end{document}